# Tuning of SiV quantum emission in nitrogen-doped nanodiamonds by dual-color excitation

A.A. Zhivopistsev[1,+], A. M. Romshin[1,+,*], A. V. Gritsienko[3,4], D.G. Pasternak[1], R.K. Bagramov[2], V.P. Filonenko[2], F. M. Maksimov[3,5], A. I. Chernov[3,5], A. M. Skomorokhov[6], A.A. Rodionov[7], N. I. Kargin[8], I. I. Vlasov[1]

*1- Prokhorov General Physics Institute of the Russian Academy of Sciences, 119991 Moscow, Russia*
*2- Vereshchagin Institute of High-Pressure Physics RAS, Troitsk, Moscow Region, Russia*
*3 - Moscow Institute of Physics and Technology (MIPT), Dolgoprudny, Russia*
*4 - P. N. Lebedev Physical Institute of the Russian Academy of Sciences, 53 Leninskiy Pr., Moscow 119991, Russia*
*5 - Russian Quantum Center, 30, Bolshoy Bulvar, Building 1, Skolkovo Innovation Center, Moscow, Russia*
*6 - Ioffe Institute, St. Petersburg 194021, Russia*
*7 - Institute of Physics, Kazan Federal University, 18 Kremlevskaya Street, 420008 Kazan, Russia*
*8 - National Research Nuclear University MEPhI, 31 Kashirskoe sh., Moscow, 115409, Russia*

[+] - these authors contributed equally, [*] - corresponding author

## Abstract

The charge dynamics of silicon–vacancy (SiV) centers have been investigated for the first time in high-pressure high-temperature nanodiamonds (NDs) with varying concentrations of substitutional nitrogen ($N_S$). We demonstrate a controlled sixfold enhancement of $SiV^–$ photoluminescence (PL) under dual-color excitation, consisting of strong red (~660 nm) illumination combined with weak green (~530 nm) excitation. The measured dependencies of $SiV^–$ PL lifetime and intensity on excitation wavelength, together with the enhancement dependence on $N_S$ concentration in the studied nanodiamonds, provide unambiguous evidence of the involvement of donor nitrogen in $SiV^–$ emission dynamics. Saturation curves and second-order PL intensity correlation measurements further indicate suppression of the population of the optically inactive $SiV^{2–}$ state upon the addition of green excitation. These results unlock a practical pathway toward engineering optically-controlled and scalable quantum emitters based on SiV-luminescent diamond nanoparticles.

Silicon–vacancy centers in diamond represent one of the most promising platforms for the implementation of quantum photonic devices [1]. Due to their unique combination of a narrow and bright zero-phonon line (ZPL), photostability, and inversion symmetry that ensures resilience to external electric fields, SiV centers are particularly attractive for quantum optics applications, including single-photon sources [2,3], quantum memory [4,5], and optical sensors [6,7].

Nevertheless, the practical use of SiV centers is significantly limited by the need for additional donor-type defects in the diamond lattice to stabilize the negatively charged state of the center. It is well established that, depending on the optical excitation conditions and the local electronic environment, SiV centers can switch between three stable charge configurations [8,9]: neutral ($SiV^0$), negatively charged ($SiV^–$), and doubly negatively charged ($SiV^{2–}$). The $SiV^–$ state exhibits bright fluorescence at 737 nm even at room temperature, whereas the $SiV^0$ state, emitting at 946 nm, can be observed only at cryogenic temperatures. Since no luminescence has been reported for the $SiV^{2-}$ configuration, it is generally regarded as an optically dark state.

Charge conversion of $SiV^-$ centers can proceed via several mechanisms. When the photon energy is sufficient to ionize an electron from $SiV^-$ into the conduction band, the neutral state $SiV^0$ can be formed. Recent studies have reported a decrease in the intensity of the $SiV^-$ ZPL in bulk diamond crystals under additional ultraviolet (UV) [10] or blue-light illumination [11]. Pederson et al. experimentally demonstrated that sub-millisecond deep-UV pulses induce the neutralization of $SiV^-$ to $SiV^0$ [10]. Interestingly, the application of a static electric field also reduces the $SiV^-$ ZPL brightness [12,13] , but in this case through recombination of the center's hole state with a valence-band electron, resulting in conversion into the optically dark $SiV^{2-}$ state. It should be noted that direct charge conversion of $SiV^-$ centers has been predominantly observed in high-purity crystals with substitutional nitrogen ($N_S$) concentrations below 5 ppb. Another pathway involves sequential charge conversion of $SiV^-$ centers mediated by auxiliary defects in the diamond lattice, such as $N_S$ [14–17]. In recent years, charge-conversion mechanisms of SiV centers have been extensively investigated using dual-color optical microscopy, in which green light efficiently ionizes $N_S$, while red light probes the charge state of SiV near the $N_S$ ionization threshold, both in bulk CVD diamonds with low nitrogen doping [17] and in samples with $N_S$ concentrations of several ppm [14–16]. Despite the critical role of substitutional nitrogen in SiV charge conversion, the dependence of SiV luminescence on the concentration of this photoactive donor impurity has not yet been systematically studied, either in bulk or nanodiamonds.

This study presents a detailed investigation of the PL behavior of $SiV^-$ centers in nanodiamonds synthesized by the high-pressure high-temperature (HPHT) method with controlled substitutional nitrogen content. For the first time, we applied a dual-color optical excitation technique to nanodiamonds in which no diffusion of free charge carriers (if generated) occurs beyond the excitation region.

A series of nanodiamond samples with varying concentrations of substitutional nitrogen was synthesized for this study. The nitrogen concentration was controlled by adjusting the adamantane/DND weight ratios to 2:1 (500 ppm), 10:1 (137 ppm), 25:1 (59 ppm), 100:1 (17 ppm), 300:1 (7 ppm), 1000:1 (3.5 ppm), and 10000:1 (2 ppm) in the precursor mixture (see Supplementary Information [18]; see also reference [19] therein). In what follows, the sample names correspond to these adamantane/DND ratios. Preliminary characterization of the samples was carried out using luminescence spectroscopy. Representative spectra were recorded under 473 nm excitation from micrometer-sized ND aggregates, as shown in **Fig. S2** (see Supplementary Information [18]). For the 2:1 and 10:1 samples, the PL-spectra exhibit characteristic ZPLs of H3 centers at 504 nm, as well as $NV^0$ and $NV^-$ centers at 575 nm and 638 nm, respectively, accompanied by broad phonon sidebands. In contrast, no luminescence from these defects was detected for samples with ratios exceeding 100:1. Across all samples, however, strong emission from $SiV^-$ centers was consistently observed, featuring a sharp ZPL at 738 nm with a full width at half maximum (FWHM) of approximately 5 nm.

We began the investigation of charge conversion under different optical excitation parameters with the 300:1 sample, since its luminescence spectrum showed no emission from other nitrogen-related defects. Individual nanodiamond particles were selected, with post-verification of their sizes (200 nm to 1 μm) performed using scanning electron microscopy (**Fig. S3**). In the first step, luminescence maps of SiV center ensembles were acquired under standalone 532 nm (green excitation, GE) and 660 nm (red excitation, RE) excitations at comparable power of 10 mW (**Fig. 1a**). We found that the luminescence intensity under GE was consistently higher than under RE. The SiV luminescence exhibited saturation behavior described by the standard dependence $I(P) = P \cdot I^{\infty}/(P + P_{sat}) + k \cdot P$, where $I^{\infty}$ is the peak count rate in the limit of large $P$, $P_{sat}$ is the saturation power, and $k$ is the background coefficient. For ND1 on **Fig. 1**, the saturation intensity under RE was $I^{\infty}_{RE} = 421$ kcounts/s, approximately two times lower than under GE. The photon count difference between individual nanodiamonds varied from 1.5 to 2.5 times, with an average GE-to-RE enhancement of about 2.1. At the same time, the ensemble of SiV centers reached saturation 2.3 times faster under RE than under GE, with $P^{RE}_{sat} = 8.9$ mW and $P^{GE}_{sat} = 20.5$ mW. A similar effect was previously reported in Ref. [20] and attributed to the broad resonance band of PL excitation, arising from the transition of the $SiV^-$ center from the ground state $^2E_g$ to the excited state $^2A_{2u}$, , in addition to the central transition into the radiative state $^2E_u$. Excitation in the green spectral region (~530 nm) overlaps with a broad (~100 nm) absorption band of the SiV center, which originates from the $^2A_{2u}$ state lying deep within the diamond valence band and strongly mixed with its electronic states. This mixing explains the higher excitation efficiency.

When nanodiamonds were subjected to strong red excitation (RE, ~10 mW), the addition of weak green excitation (GE, on the order of 10–100 μW) led to an unexpected and significant increase in luminescence intensity. Under such combined excitation (CRGE), the observed SiV PL substantially exceeded not only the response to green or red excitation individually but also their sum, indicating that the effect is not a simple superposition of excitations but rather a more complex interaction between the quantum states of the color centers and the external optical field. Assuming that the luminescence enhancement under CRGE depends on the contributions of both red and green excitation rates individually, we investigated the dynamic response of SiV centers to incremental increases in GE power following the protocol illustrated in **Fig. 1a**. The selected ND1 was continuously illuminated with RE, while GE was applied intermittently for short intervals of about 10 s. The PL-contribution from GE alone was measured separately (**Fig. 1c**). From the recorded time traces, absolute photon count rates under RE ($I_{RE}$) and CRGE ($I_{CRGE}$)were extracted, and their difference was calculated as $\Delta I = I_{CRGE} - I_{RE}$. The dependencies of these parameters on optical power in GE, RE, and CRGE were then established (Fig. 1c–d). The admixture of 0.4 mW green light to RE resulted in a substantial increase in saturation intensity, reaching up to six times the value observed under standalone RE. It was also found that the SiV luminescence enhancement under CRGE occurs even at lower green powers, with the relative increase compared to $I_{RE}$ depending on the primary excitation power in the CRGE regime. To quantify this effect, we introduce an enhancement factor $\eta = \frac{\Delta I}{I_{RE}}$, which represents the relative increase in count rate under CRGE $\Delta I$ with respect to RE alone $I_{RE}$.

In order to identify the most efficient excitation conditions for SiV centers, PL-measurements as a function of the GE power were carried out at different RE power levels ranging from 0.2 mW to 10 mW (**Fig. 1d**). We found that the higher the RE photoexcitation rate, the less GE power was required to achieve effective SiV PL. The data were fitted using a modified saturation function shifted by a constant $I_{RE}$, representing the photon count rate of SiV emission under standalone RE at a given power. From this analysis the phenomenological parameters $P_{sat}$ and $\Delta I^{\infty}$ were extracted. The results in **Fig. 1d** show that the SiV luminescence intensity increases significantly with RE power and exhibits saturation behavior at higher power levels. **Figure 1e** further demonstrates that $P_{sat}$ decreases as $\Delta I^{\infty}$ increases, indicating more efficient energy transfer under combined excitation. Moreover, the enhancement of SiV luminescence was found to increase incrementally with RE power, reaching a maximum at approximately 10 mW, beyond which the enhancement rate saturates. This suggests that although higher RE power improves luminescence, the efficiency gain diminishes above a certain threshold.

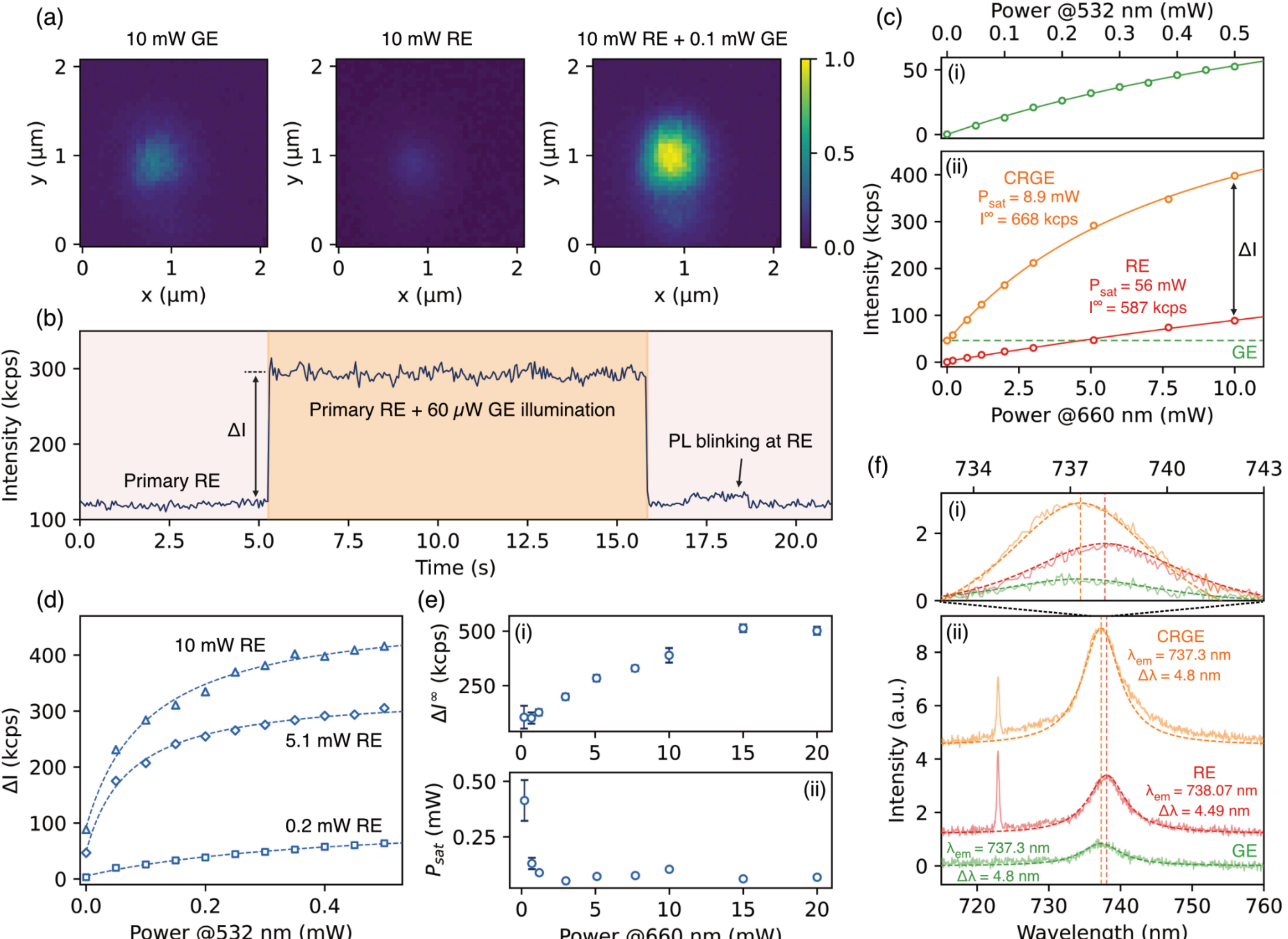

**Fig. 1**. PL of SiV centers in a pre-selected ND (300:1 sample) under different excitations. (a) PL-maps recorded at RE and GE (10 mW each) and CRGE (10 mW at 660 nm, 100 μW at 532 nm). (b) Time traces of the photon count rate detected in the 720-750 nm bandwidth under 9.1 mW RE (red shadowed regions) and CRGE (9.1 mW at 660 nm, 60 μW at 532 nm; orange shadowed regions). (c) Saturation curves under GE (inset (i)), RE, CRGE with 400 μW 532 nm (inset (ii)). (d) Photon count rate shifts $\Delta I$ measured at various GE power levels under combined excitation with three fixed RE powers: 0.2 mW (squares), 5.1 mW (diamonds), and 10 mW (triangles). (e) RE power dependence of saturation parameters: (i) $\Delta I^{\infty}$ and (ii) $P_{sat}$, extracted from the curves shown in (d). (f) PL-spectra of SiV centers recorded under 1.6 mW RE, 400 μW GE, and CRGE (9.1 mW at 660 nm, 10 μW at 532 nm). Inset (i): statistical distribution of the ZPL spectral position across 15 NDs under the same RE and CRGE powers. Markers and solid lines: RE - red, GE - green, CRGE - orange.

The PL of $SiV^-$ centers can be described within a three-level system formalism, where the two energetically separated levels correspond to the SiV ZPL (transition rate $k_{21}$), and the third, intermediate level acts as a shelving or trapping state [2]. This state is empirically associated with nonradiative charge transfer processes characterized by the rate constant $k_{23}$, which reduce the overall radiative efficiency of the centers. Following the formalism in Ref.[21] and solving the rate equations for the level populations, the saturation of SiV PL can be expressed in terms of the transition rates as $I^{\infty}(k_{23}, k_{31}) = \frac{k_{21}}{(1+k_{23}/k_{31})}$ and $P_{sat}(k_{23}, k_{31}) = \frac{k_{21}+k_{23}}{\sigma}/(1 + k_{23}/k_{31})$, where $\sigma = k_{12}/P$ is the excitation constant depending on the excitation energy and absorption cross-section. The observed enhancement of PL under CRGE suggests a dependence of the transition rates on green illumination. In particular, the capture rate $k_{23}$ and the relaxation rate $k_{31}$ govern this dynamics and are functions of the green excitation power $P_{GE}$. A decrease in $k_{23}$ corresponds to the suppression of $SiV^{-}$ capture into the shelving state, whereas an increase in $k_{31}$ corresponds to de-shelving of |3>. While both processes may occur, we consider suppression of the |2>→|3> transition the more probable mechanism. We attribute the origin of the shelving state to substitutional nitrogen $N_S$, which, acting as an electron donor, can ionize $SiV^{-}$ into the optically dark $SiV^{2-}$ state. Under CRGE conditions, the reduction of $k_{23}(P_{GE})$ is explained by blocking the |2>→|3> transition due to green-light-induced ionization to conduction band $N^0$→$N^+$[22,23]. According to this model, the observed dependence of $SiV^{-}$ PL on $P_{GE}$ can be approximated as $I(P_{GE}, P_{RE}) = \frac{k_{21}}{1+(k_{21}+k_{23}(P_{GE},P_{RE}))/(\sigma_{RE}P_{RE}+\sigma_{GE}P_{GE}) +k_{23}(P_{GE},P_{RE})/k_{31}}$. In our interpretation, the variation of $k_{23}$ with green illumination is ascribed to the photoionization $N^0$→$N^+$, which prevents electron transfer to $SiV^{-}$ and thereby suppresses the |2>→|3> transition. The efficiency of this process increases with $P_{GE}$. At the same time, the experimentally observed increase in saturation intensity with higher RE power is attributed to a reduced population of $k_{23}$ the ground state with rising $P_{RE}$. Taking these mechanisms into account, the dependence of $k_{23}$ can be expressed in factorized form $k_{23} = k_{23}^{(0)} \cdot f_{GE}(P_{GE}) \cdot f_{RE}(P_{RE})$, where $k_{23}^{(0)}$ is the capture rate without illumination, $f_{GE}(P_{GE}) = \frac{1}{1+P_{GE}/P^0_{N_s}}$ describes suppression of capture due to donor ionization by GE ($P^0_{N_s}$ is the power at which half of the nitrogen donors are ionized), and $f_{RE}(P_{RE}) = 1 + \frac{P_{RE}}{P^*_{RE}}$ accounts for the enhanced capture with increasing RE power due to the redistribution of population from |1>.

The power-dependent variation of the recombination rate of $SiV^{-}$ with $N_S$ is expected to manifest in the PL behavior of NDs containing a discrete number of SiV centers. Analysis of the second-order correlation function $g^{(2)}(\tau)$ for one such ND revealed that the depth of the antibunching dip at $\tau = 0$, which serves as an indicator of the number of independent emitters and is given by $g^{(2)}(0) = (1 - 1/n)$ (where $n$ is the number of emitters), remained nearly unchanged under RE, GE, and CRGE excitation conditions (**Fig. 2a**) and corresponded to $n = 12$. Minor fluctuations may originate from inaccuracies in accounting for the background signal across different power modes. In contrast, under red excitation alone, with powers of several milliwatts, the correlation function $g^{(2)}(\tau)$ exhibited pronounced photon bunching on timescales of 15–25 ns. This effect is attributed to luminescence intermittency of SiV centers on these timescales. Upon addition of green illumination (10 μW to 1–5 mW) in the CRGE regime, the bunching amplitude was significantly reduced. Within the three-level system formalism, $g^{(2)}(\tau)$ can be expressed as $g^{(2)}(\tau) = 1 - (1 + a) \cdot e^{-\tau/\tau_1} + a \cdot e^{-\tau/\tau_2}$, where $\tau_1$ stands for antibunching, while $a$ and $\tau_2$ describe bunching. Since the antibunching contribution of SiV centers is relevant only on sub-nanosecond timescales, the expression simplifies for longer dynamics to $g^{(2)}(\tau) \approx 1 + a \cdot e^{-\tau/\tau_2}$. In terms of transition rates, the bunching parameters are expressed as $a = \frac{k_{12} \cdot k_{23}(P_{GE})}{k_{31} \cdot (k_{12} + k_{21})}$ and $\tau_2 = \left(k_{31} + \frac{k_{12} \cdot k_{23}(P_{GE})}{k_{12} + k_{21}}\right)^{-1}$. Analysis of their dependence on green excitation power showed that with increasing $P_{GE}$, the bunching amplitude $a$ gradually decreased, reaching a plateau at approximately 0.5 mW (**Fig. 2b**). This behavior is consistent with a model in which green illumination suppresses capture into the shelving state via ionization of $N^0$ donors, thereby reducing $k_{23}(P_{GE})$. Consequently, the suppression of bunching reflects a reduced probability of capture into the dark state, while the return time from this state remains unaffected.

When analyzing the luminescence spectra of $SiV^-$ ensembles in nanodiamonds with a nitrogen concentration of 4 ppm under different excitation conditions, we observed an unusual blue shift of the ZPL under CRGE and GE compared to RE (**Fig. 1e**). The magnitude of this shift varied from 0.3 to 0.8 nm between particles. At the same time, the ZPL linewidth reproducibly increased by up to 10%. Since no such shifts were observed in particles with lower adamantane/DND ratios, the ZPL modifications were interpreted as a local influence of substitutional nitrogen on the electronic structure of SiV centers. We propose that the emergence of positively charged $N^+$ centers in the vicinity of $SiV^-$ leads to a quadratic Stark effect, which shifts the ZPL toward shorter wavelengths. Moreover, a contribution from local lattice compression around the SiV center induced by ionized nitrogen cannot be excluded. The ZPL broadening may be attributed to statistical variations in the orientation and distance of ionized donors, causing fluctuations of the local potential and spectral diffusion, similar to that reported previously for NV centers[24]. A detailed study of the relative contributions of charge-related and strain-induced effects to the ZPL shift and broadening in nanodiamonds with different $N_S$ concentrations will be presented elsewhere.

Interestingly, gradual cooling to cryogenic temperatures (~10 K) enhanced the role of green illumination in the SiV PL. As shown in **Fig. 3a**, the enhancement factor increased by approximately twelvefold upon cooling, indicating a pronounced strengthening of the combined excitation effect at cryogenic conditions. Notably, this increase in the enhancement factor occurred alongside a rise in SiV PL under RE excitation with decreasing temperature. We therefore conclude that the capture of a donor electron by the $SiV^-$ center proceeds more efficiently under conditions where phonon interactions are suppressed.

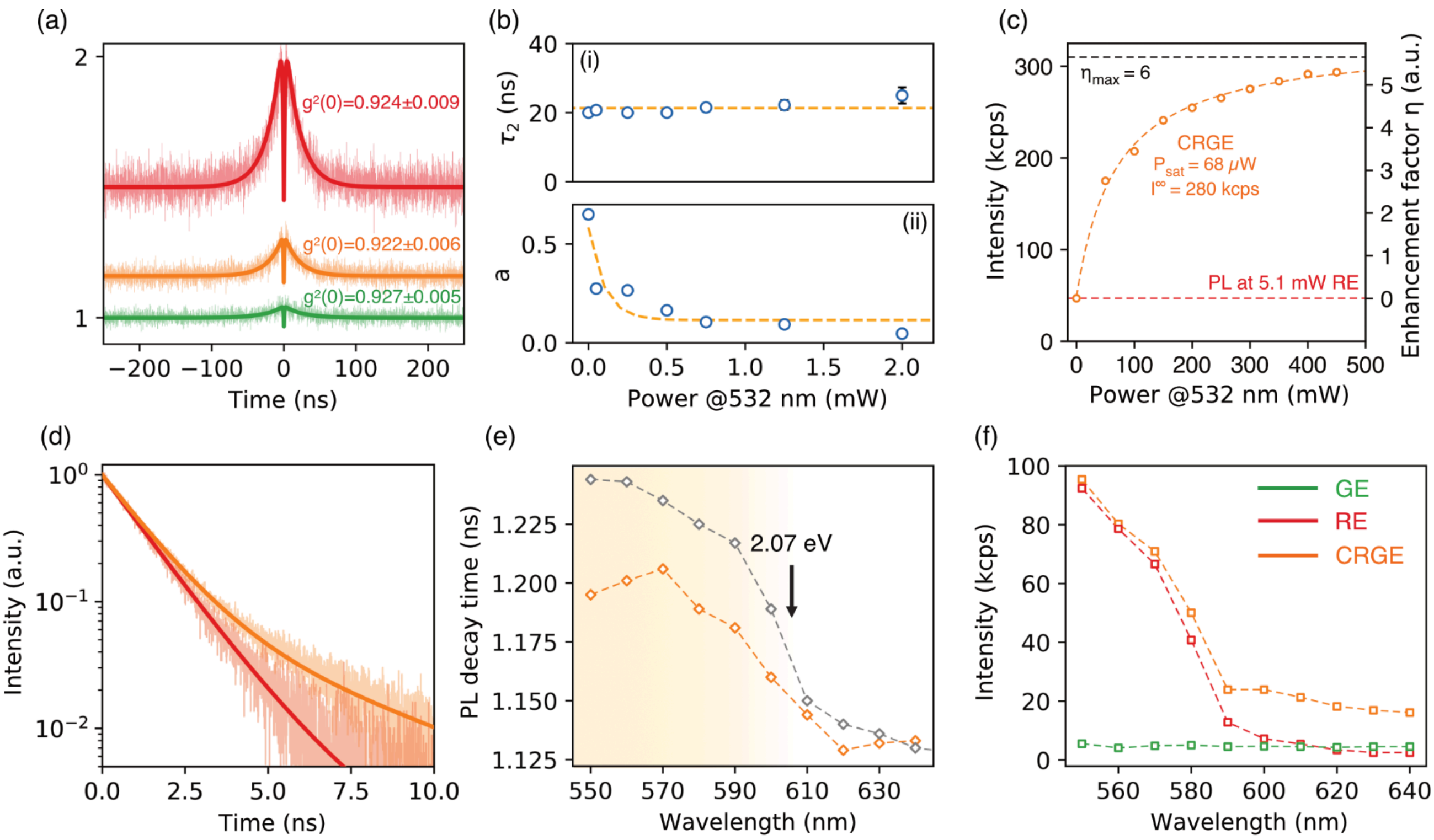

**Fig. 2.** PL of discrete SiV emitters in NDs. (a) Second order correlation function measured for ND with discrete emitters under RE (10 mW, red), GE (10 mW, green), and CRGE (10 mW RE, 200 μW GE, orange). The antibunching dip at $\tau = 0$ remains nearly constant. (b) The dependence of (i) bunching amplitude and (ii) bunching characteristic timescale on GE power. (c) Photon count rate increase $\Delta I^{\infty}$ measured at various GE power levels under CRGE (5.1 mW at 660 nm) and the corresponding changes in the enhancement factor η (right axis). PL spectra of SiV centers recorded under RE (5.5 mW) and CRGE (5.5 mW at 660 nm and 100 μW at 532 nm). (d) Fluorescence decay curves obtained under pulsed RE at 640 nm standalone (red) and under synchronized pulsed GE (orange). (e) Fluorescence decay times obtained under standalone spectrally tunable pulsed excitation (550–640 nm, average power 50 μW, gray) and under synchronized pulsed RE at 640 nm (average power 10 μW, orange). (f) Photon count rates of SiV PL under the same excitation conditions as in (e). Green dashed lines and scattered points correspond to standalone pulsed GE.

Next, time-resolved measurements of SiV luminescence were performed using a wavelength-tunable pulsed laser synchronized with a red pulsed laser (640 nm). The average power of the red laser was 60 μW, while the tunable laser varied between 10 μW and 50 μW. The fluorescence decay times of SiV centers were measured under two conditions: (1) tunable excitation with an average power of 50 μW, and (2) tunable excitation with an average power of 10 μW synchronized with RE. It was observed that the fluorescence lifetime of SiV centers decreased by approximately 8–10%, with a noticeable inflection point around 600 nm (2.07 eV) in both cases, but more pronounced under condition (1) with stronger tunable excitation. In the short-wavelength region, the increase in fluorescence lifetime can be explained by the reduced contribution of nonradiative transitions $\tau = \frac{1}{k_r + k_{nr}}$, where $k_r$ and $k_{nr}$ are the radiative and nonradiative decay rates, respectively. The photon count rate showed a concomitant decrease for both excitation cases, as presented in **Fig. 2f**. Notably, the SiV PL intensity dropped significantly with increasing excitation wavelength, reaching a minimum plateau again at 600 nm. The PL level of SiV centers in the presence of green illumination was found to be five times higher than under standalone tunable excitation, and 2.8 times greater than the sum of photon counts from the two standalone excitations.

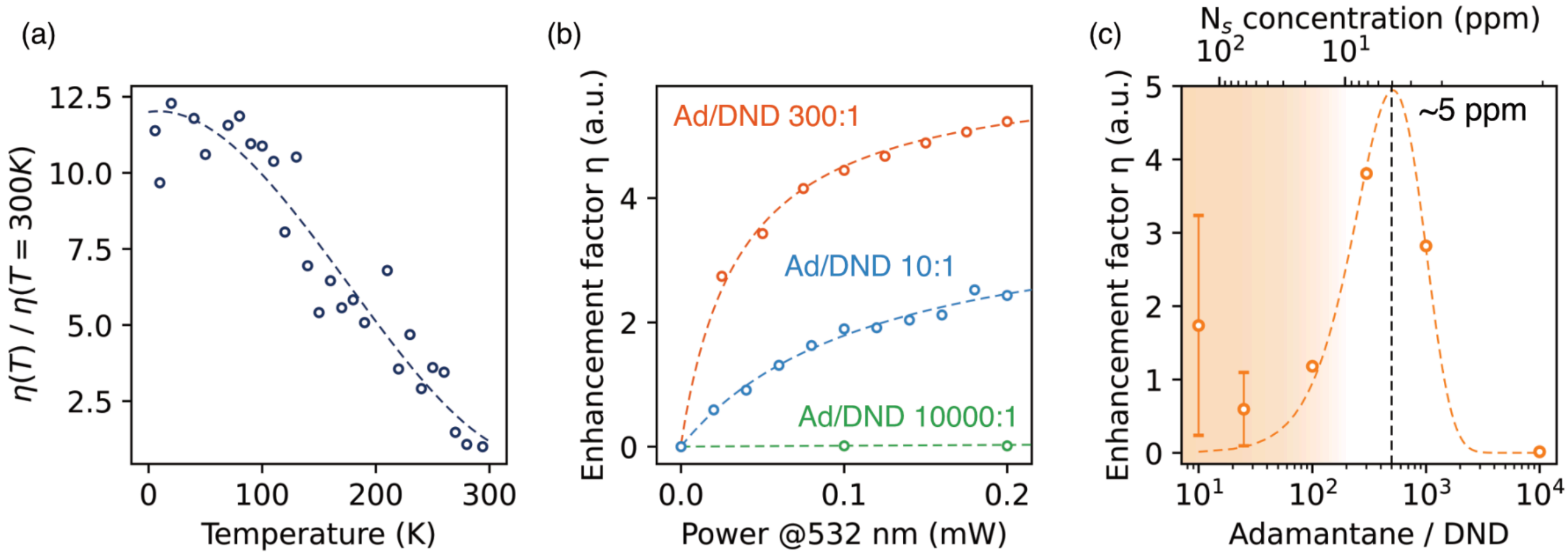

**Fig. 3.** (a) Temperature dependence of relative η-factor (normalized by η-factor at room temperature); (b) Saturation curves for nanodiamonds with adamantane/DND ratio 10:1 (blue), 300:1 (orange), 10000:1 (green); (c) Enhancement factor dependence on adamantane/DND ratio or nitrogen concentration.

Finally, the effect of combined excitation on the PL of SiV centers was studied for samples with different concentrations of $N_S$. A pronounced dependence on nitrogen content was observed: for samples with low nitrogen concentration, the influence of GE was almost negligible (**Fig. 3b–c**), with the enhancement factor η not exceeding 0.05–0.1. As the fraction of $N_S$ increased the enhancement factor η rose to 6–7 due to the greater number of nitrogen donors available for recombination with $SiV^-$. However, at higher nitrogen concentrations (>20 ppm), η decreased to ~0.5, which we attribute to the large number of donors in close proximity to $SiV^-$. In this regime, weak green illumination is insufficient to efficiently ionize nitrogen, leading to a substantial suppression of the $SiV^-$ enhancement effect.

Notably, the position of the enhancement maximum as a function of $N_S$ concentration correlates well with values at which efficient donor-assisted charge conversion in CVD diamonds has been reported in previous studies [16]. The observed maximum at ~5 ppm indicates the existence of an optimal balance between donor concentration and the probability of effective charge conversion. According to the nearest-neighbor probability distribution for 5 ppm $N^0$ , the average distance from a SiV center in the studied sample to its two nearest nitrogen atoms is estimated to be ~6 nm. However, this estimate should be regarded as approximate, since it does not account for the inhomogeneous distribution of nitrogen impurities across different diamond growth sectors.

One of the key findings of our study is the observation of a low threshold GE power density required to induce the enhancement of $SiV^-$ photoluminescence: values below 10 $W/mm^2$ (corresponding to an absolute power of ~10 μW) already produce a pronounced enhancement, indicating highly efficient charge conversion. This sensitivity is consistent with the results of Ref. [25], where saturation of optical hyperpolarization of nitrogen donors was achieved at intensities of ~10 $mW/mm^2$ and attributed to the ionization of $N^0 \rightarrow N^+$ centers in the vicinity of NV centers. The comparable orders of magnitude and excitation conditions suggest a common origin of the underlying processes associated with photoinduced nitrogen ionization.

Experiments with the tunable excitation source showed that the charge conversion $N^0 \rightarrow N^+$ becomes efficient above a photon energy threshold of >2–2.1 eV. This value is consistent with the ionization threshold of substitutional nitrogen previously determined from absorption spectra [26].

In conclusion, we have for the first time investigated the charge dynamics of $SiV^-$ centers in nanodiamonds with varying concentrations of substitutional nitrogen. We observed a sixfold increase in the $SiV^-$ photoluminescence intensity under dual-wavelength excitation, consisting of strong red (~660 nm) illumination combined with weak green (~530 nm) excitation. The inflection point at 2.07 eV in the excitation-wavelength dependence of both the fluorescence lifetime and intensity, together with the dependence of the enhancement on $N_S$ concentration in the studied nanodiamonds, provides unambiguous evidence for the involvement of donor nitrogen in the emission dynamics of $SiV^-$. Saturation curves and second-order intensity correlation measurements further indicate suppression of the population of the metastable $SiV^{2-}$ state upon the addition of green excitation.

The overlap between the electronic structure of $SiV^-$ centers and the levels of substitutional nitrogen opens new avenues for studying coherent interactions between $SiV^-$ and $N_S$. A particularly relevant direction is the investigation of charge transport from donor nitrogen to $SiV^-$ in a magnetic field, where spin–orbit degeneracy is lifted. Of special interest is the prospect of constructing a quantum cell in which nitrogen serves as a long-lived spin register, while the $SiV^-$ center provides a platform for optical initialization and readout of the $N_S$ state. Such hybrid systems could form the basis for scalable quantum memories and multi-addressable quantum nodes.

## Acknowledgements

This study was supported by Ministry of Science and Higher Education of the Russian Federation (agreement/grant 075-15- 2025-609).

# Supplementary Information

# Tuning of SiV quantum emission in nitrogen-doped nanodiamonds by dual-color excitation

A.A. Zhivopistsev[1,+], A. M. Romshin[1,+,*], A. V. Gritsienko[3,4], D.G. Pasternak[1], R.K. Bagramov[2], V.P. Filonenko[2], F. M. Maksimov[3,5], A. I. Chernov[3,5], A. M. Skomorokhov[6], A.A. Rodionov[7], N. I. Kargin[8], I. I. Vlasov[1]

*1- Prokhorov General Physics Institute of the Russian Academy of Sciences, 119991 Moscow, Russia*
*2- Vereshchagin Institute of High-Pressure Physics RAS, Troitsk, Moscow Region, Russia*
*3 - Moscow Institute of Physics and Technology (MIPT), Dolgoprudny, Russia*
*4 - P. N. Lebedev Physical Institute of the Russian Academy of Sciences, 53 Leninskiy Pr., Moscow 119991, Russia*
*5 - Russian Quantum Center, 30, Bolshoy Bulvar, Building 1, Skolkovo Innovation Center, Moscow, Russia*
*6 - Ioffe Institute, St. Petersburg 194021, Russia*
*7 - Institute of Physics, Kazan Federal University, 18 Kremlevskaya Street, 420008 Kazan, Russia*
*8 - National Research Nuclear University MEPhI, 31 Kashirskoe sh., Moscow, 115409, Russia*

[+] - these authors contributed equally, [*] - corresponding author

## Materials and methods

Nanodiamonds were produced using the high-pressure high-temperature (HPHT) method from a mixture of adamantane ($C_{10}H_{16}$, Aldrich, St. Louis, MO, USA, 99% purity) and detonation nanodiamonds (DND, Adamas Nanotechnologies Inc., Raleigh, NC, USA, average size 3–4 nm) containing ~1% nitrogen impurity. The samples were synthesized at a pressure of ~7.5 GPa and a temperature range of 1200–1500 °C for 20 s. For sample preparation, the following weight ratios of adamantane to DND were used: 2:1, 10:1, 100:1, 300:1, 1000:1, and 10000:1. In what follows, the sample names correspond to these ratios. We assume that relative variations in the N/C ratio in the precursor mixture are proportional to those in the synthesized samples. Thus, by determining the nitrogen concentration in one of the samples using electron paramagnetic resonance (EPR), the nitrogen concentration in other samples can be estimated based on the relative changes in the precursor N/C ratio. To form SiV centers in the NDs, tetrakis(trimethylsilyl)silane ($C_{12}H_{36}Si_5$, Sigma-Aldrich, St. Louis, MO, USA, >97%) was added to the initial mixture.

To measure the concentration of substitutional nitrogen in nanodiamonds using the electron paramagnetic resonance (EPR) method, diamond powder was placed in a glass capillary, and the EPR spectra were recorded at a temperature of 10 K. For the sample with the highest adamantane/DND (2:1), EPR measurements were carried out at a frequency of 94 GHz using a custom-built spectrometer developed at the A.F. Ioffe Institute in cooperation with the DOK company. For the sample with the lowest adamantane/DND weight ratio (10000:1), EPR measurements were performed at 9.8 GHz using a commercial ESP-300 spectrometer (Bruker, Billerica, Massachusetts). The $N_S$ concentration was determined from the linewidth of the corresponding EPR peak (**Fig. S1**) [1]. For the adamantane/DND weight ratio of 2:1, the measured concentration was 500 ppm. In contrast, for the sample with the lowest adamantane/DND ratio of 10000:1, the $N_S$ concentration was found to be ≈2 ppm. Using these data, other $N_S$ concentrations for the remaining samples were estimated under the assumption that all nitrogen is contained in the DND particles and mixing is homogeneous. The estimated $N_S$ concentrations for other adamantane/DND weight ratio are: 10:1 ≈137 ppm, 25:1≈59 ppm, 100:1 ≈17 ppm, 300:1 ≈7 ppm, 1000:1 ≈3.5 ppm, and 10000:1 ≈2 ppm.

The synthesized nanodiamond powders were dispersed in ethanol at a concentration of 0.1 g/L and sonicated in an ultrasonic bath for 15 min to obtain a uniform suspension. The nanodiamonds were then deposited on silicon substrates by drop-casting. The size and morphology of the crystals were examined using a scanning electron microscope (SEM).

The luminescent properties of SiV-centers in NDs were examined in a custom-made confocal microscope (**Fig. S5**), equipped with a high-aperture objective (NA=0.95) and a piezoelectric stage (piezosystem jena TRITOR 100), enabling precise 3D scanning with nanoscale resolution. Optical images of the nanodiamonds were captured by a high-resolution CMOS camera under wide-field illumination. To excite the luminescence of SiV centers, two laser sources emitting at 532 nm (Cobolt 08-DPL) and 660 nm (Laser Quantum GEM) were used both in combination and separately. Both lasers were aligned and focused into a single point. To eliminate reflected laser light, two notch filters (532±5 nm and 660±5 nm) were included in the optical path. The spectra of SiV-centers were recorded with two different spectrometers: (1) Ocean Insight QE Pro (1800 $mm^{-1}$, slit 100 μm) and (2) custom one based on a Solar LS M266 monochromator coupled with a single-photon counting module (Excelitas SPCM–AQRH–14–FC). The same avalanche detectors together with Time Tagger 20 (Swabian Instruments) correlator were employed for the measurements of saturation curves and the second-order correlation function $g^2(\tau)$ using the Hanbury-Brown-Twiss interferometer. During these types of measurements a band pass (BP) filter in the range of 730–750 nm was used to minimize the possible influence of background signal originating from other, primarily nitrogen-related, luminescent centers (such as $H_3$ and $NV^{0/-}$) and the substrate material. The optical setup also featured two lenses and a 100-micron pinhole to suppress stray light from the substrate and enhance signal-to-noise ratio.

Measurements under combined excitation were also conducted in time-resolved mode. For this we used another optical setup including a Picoquant MicroTime 200 fluorescence confocal microscope coupled to an Olympus IX71 inverted microscope. SiV-luminescence was excited with two different pulses simultaneously by a supercontinuum laser (SpectraK NKT Photonics): (1) fixed at 532 nm with about 10 μW average power and (2) tunable within the range between 540-650 nm with ~60 μW average power. The repetition rate of both laser pulses were 80 MHz and pulse duration was 50 ps, herewith the time delay between the ones was about 400 ps. For different wavelengths of tunable laser, the pulse arrival times at the sample varied, with delays of up to 0.3 ns. Signals were acquired using a single-photon avalanche photodiode (MicroPhoton Devices,

PDM series, 35 ps timing resolution) with a PicoHarp300 (PicoQuant) counting module. We used BP filter @720-760 nm to remove the unnecessary spectral regions of the signal.

The micro-PL system for low-temperature measurements comprised a Thorlabs HNL050LB continuous-wave (CW) laser with a radiation wavelength of 632.8 nm and a KLM-532/SLN-x CW laser with a wavelength of 532 nm. The sample was mounted in a Montana Instruments Cryostation S50 with optical access. The excitation beam was focused onto the sample using a 100X Mitutoyo Plan Apo NIR HR Infinity Corrected objective (NA = 0.7). The photoluminescence signal was collected and analyzed using a HORIBA Jobin Yvon iHR320 spectrometer equipped with a SynapsePlus detector.

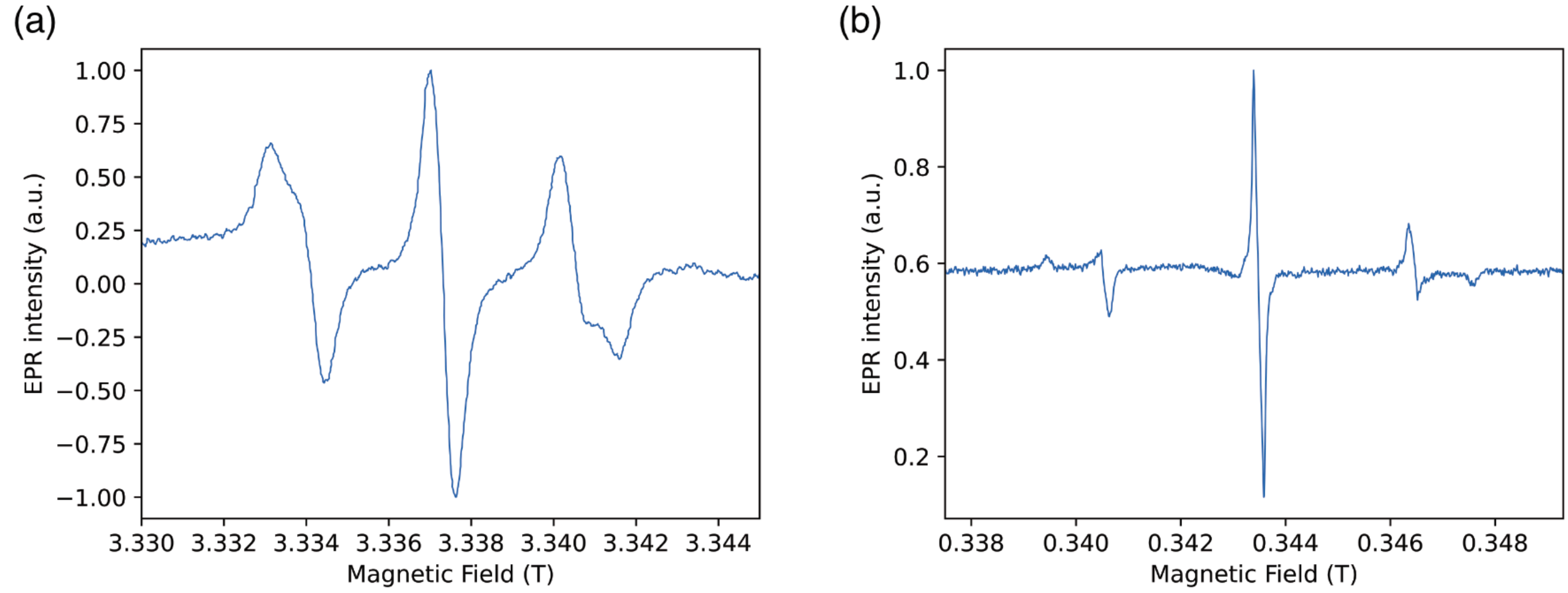

**Fig. S1**. EPR spectra of HPHT nanodiamond powder at 10 K with adamantane:DND ratios of (a) 2:1 (custom spectrometer, 94 GHz frequency) and (b) 10000:1 (9.8 GHz, Bruker ESP-300).

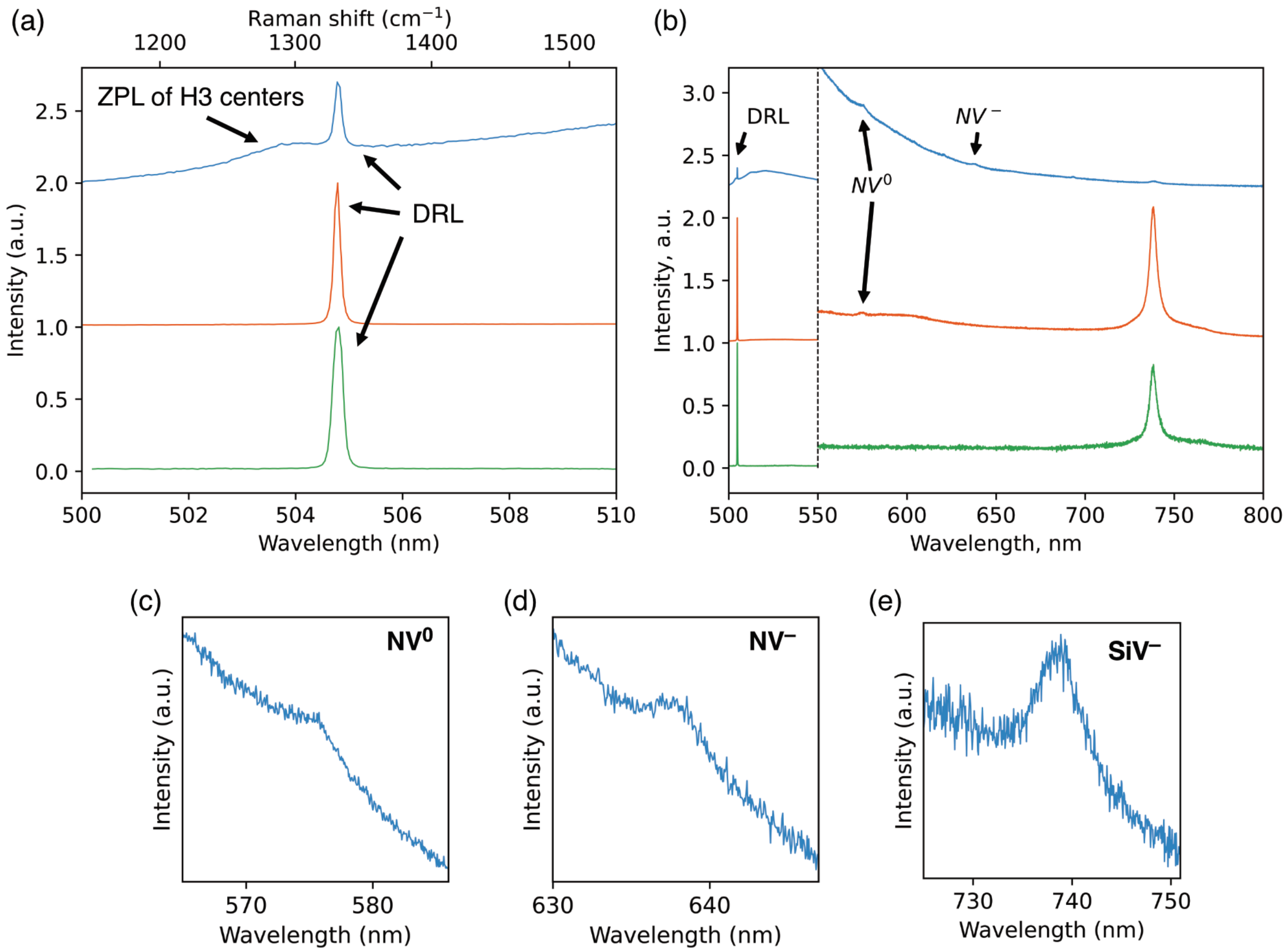

**Fig. S2**. (a) Raman and (b) PL spectra of NDs synthesized with varying adamantane/DND weight ratios: 10/1 (blue), 300/1 (orange), 1000/1 (green), under 473 nm laser excitation. All samples exhibit a pronounced and sharp diamond Raman line (DRL) at 1332.5 см$^{-1}$ (504.8 nm). The PL spectra show characteristic ZPLs of H3 centers at ~504 nm, as well as ZPLs of NV$^0$ and NV$^-$ centers at 575 nm and 638 nm, respectively, accompanied by broad phonon sidebands. The SiV$^-$ ZPL at 738 nm has a full width at half maximum approximately 5 nm.

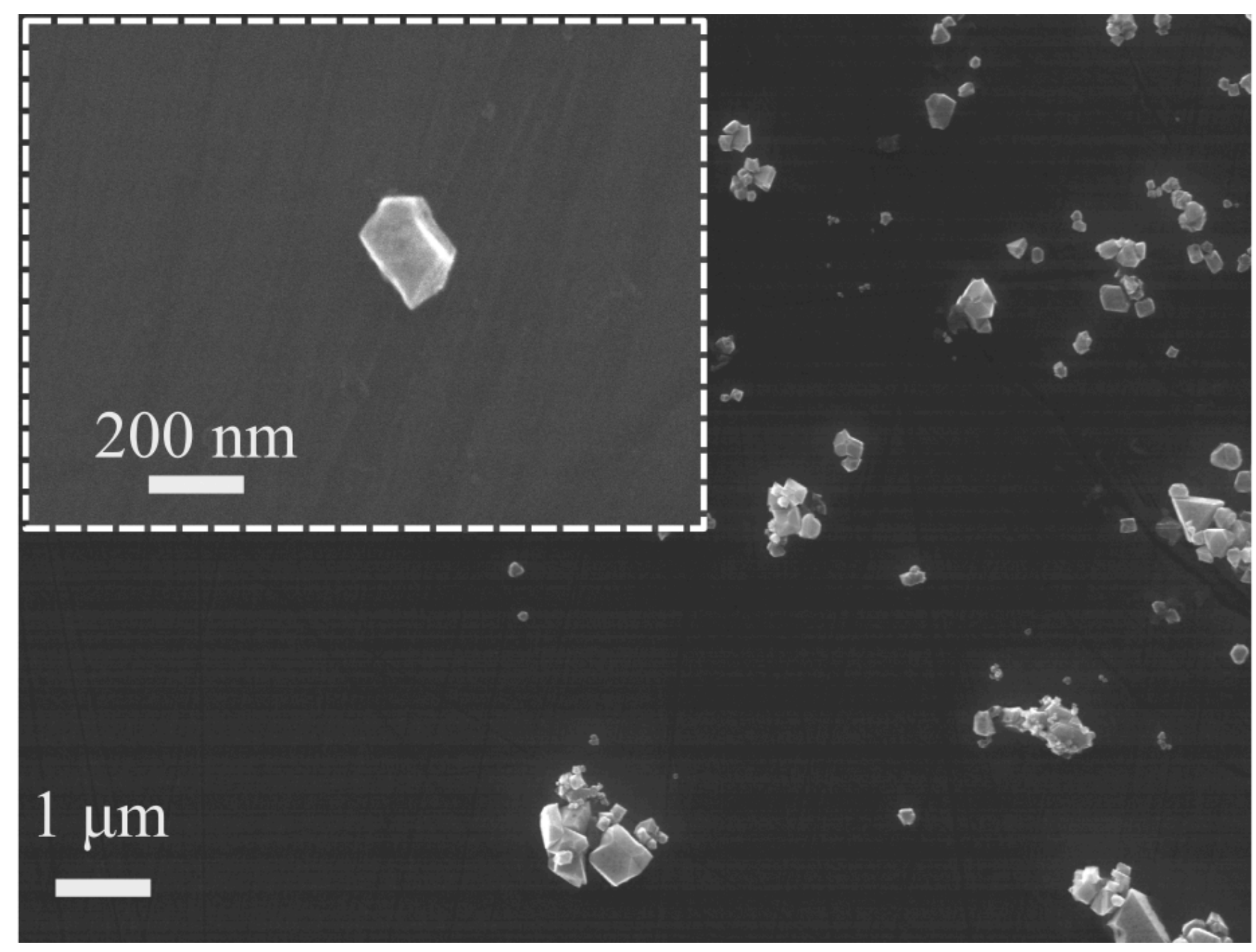

**Fig. S3**. SEM-image of HPHT NDs dispersed on the surface of silicon substrate.

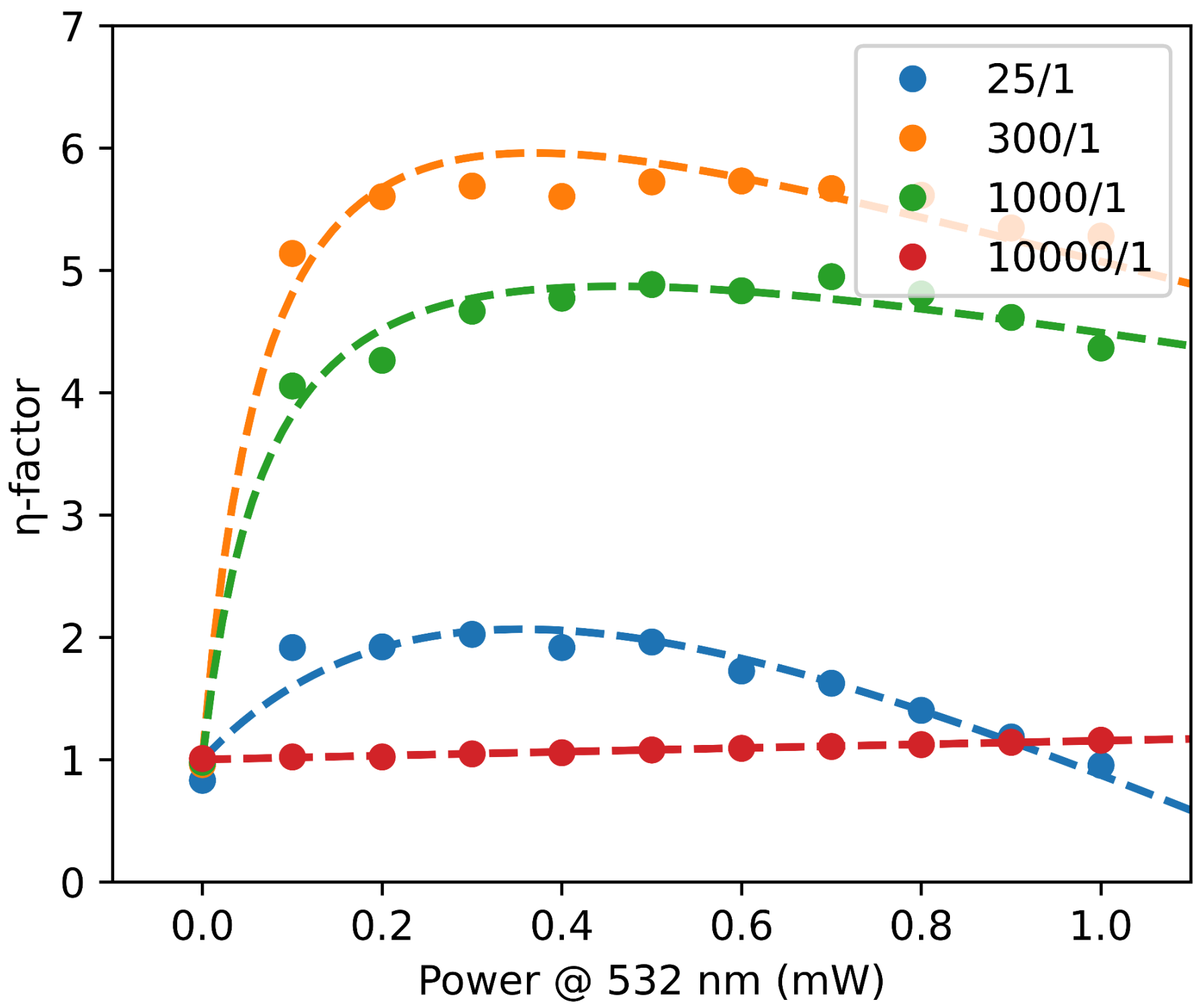

**Fig. S4**. Enhancement factor dependence on GE power obtained in CRGE (5 mW at 660 nm) for different adamantane/DND ratio: 25/1 (blue), 300/1 (orange), 1000/1 (green), 10000/1 (red).

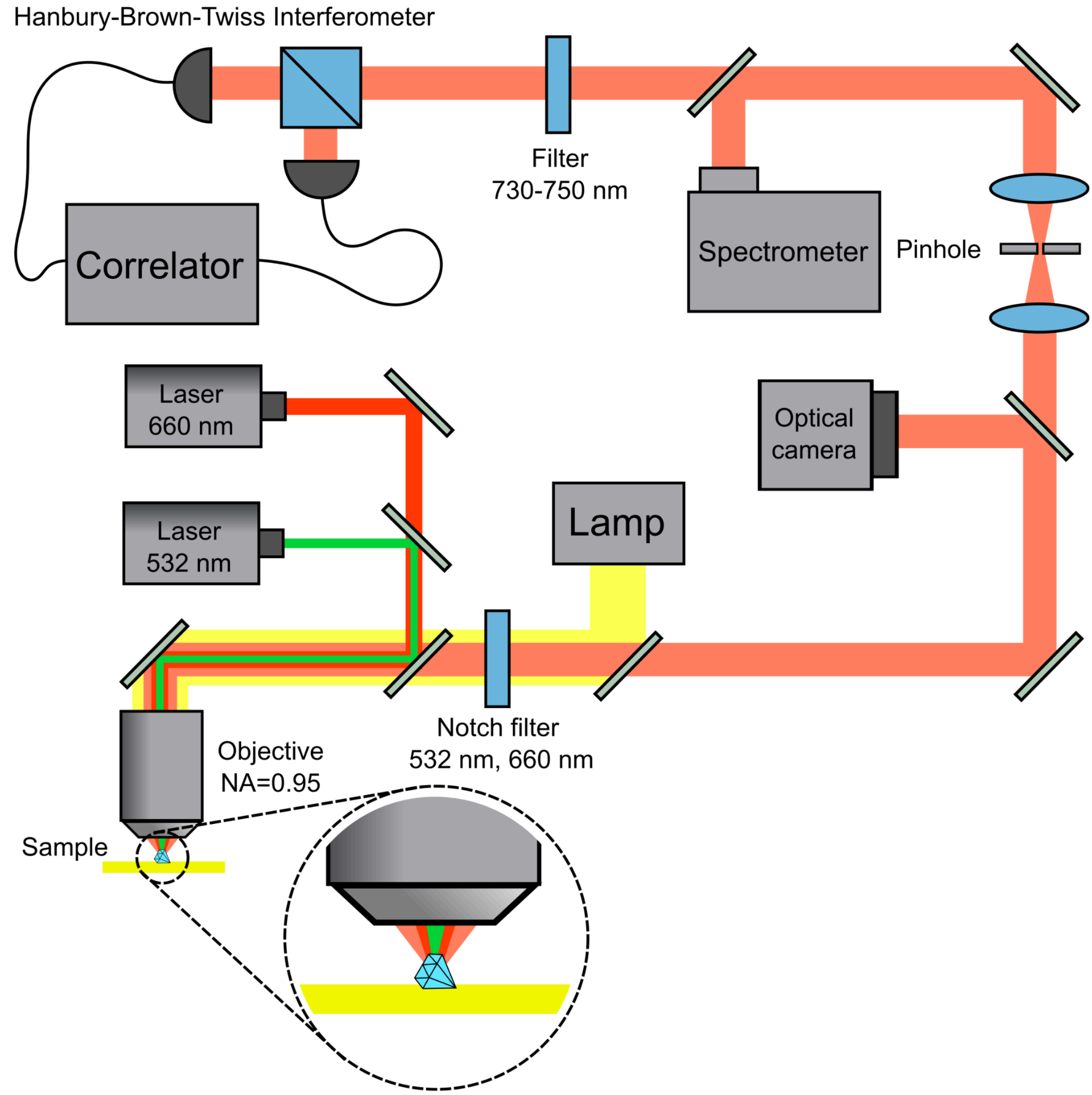

**Fig. S5**. A schematic illustration of an experimental setup. Confocal microscope with two lasers 532 nm and 660 nm, objective x100, spectrometer and Hanbury Brown & Twiss interferometer.